\documentclass[twocolumn, prper,superscriptaddress]{revtex4}%
\usepackage{amsfonts}
\usepackage{amsmath}
\usepackage{amssymb}
\usepackage{graphicx}
\usepackage{subfigure}
\usepackage{epstopdf}
\usepackage{multirow}%
\usepackage{hyperref}
\usepackage{appendix}

\hypersetup{hidelinks}
\hypersetup{
	colorlinks=true,
	linkcolor=blue,
	filecolor=blue,
	urlcolor=blue,
	citecolor=blue,
}
\providecommand{\U}[1]{\protect\rule{.1in}{.1in}}

\begin{document}
	
	\title{Rogue waves collision under incident momentum modulation in two-component Bose-Einstein condensates }
	
	\author{Zhihao Zhang}
	\affiliation{Department of Physics, Xiangtan University, Xiangtan 411100, China}

    \author{Tiantian Li}
	\email{ttli@xtu.edu.cn}
	\affiliation{Department of Physics, Xiangtan University, Xiangtan 411100, China}
 
	\author{Xiao-Dong Bai}
	\affiliation{College of Physics, Hebei Normal University, Shijiazhuang 050024, China}	

    \author{Yunbo Zhang}
	\affiliation{Department of Physics, Zhejiang Sci-Tech University, Hangzhou, 310018, China}		
	
	\author{Denglong Wang}
	\email{dlwang@xtu.edu.cn}
	\affiliation{Department of Physics, Xiangtan University, Xiangtan 411100, China}

	\begin{abstract}
	The collision dynamics of two first-order rogue waves (RWs) with opposite incident momentum in two-component Bose-Einstein condensates (BECs) is studied by solving the two-component one-dimensional Gross-Pitaevskii (GP) equation. It is demonstrated that the introduction of appropriate incident momentum successfully promotes the generation of second-order RWs in the case of relatively weaker interspecies interactions compared to intraspecific interactions. The range of incident momentum that can facilitate the generation of second-order RWs under different interspecies interaction strengths is determined, and machine learning is employed to find and analyze relationships among the interspecies interaction, the incident momentum, and the offset that can lead to the generation of second-order RWs. It shows that any two parameters above exhibit a positive or negative correlation when the third parameter is fixed. These findings provide additional possibilities for generating and controlling high-order RWs.  
		
	\end{abstract}

	\maketitle

	\section{Introduction}
 
     The realization of Bose-Einstein condensates (BECs) in weakly interacting atomic gases provides the opportunity for researchers to study the nonlinear characteristics of atomic matter waves \cite{BEC1, BEC2}. It is well established that its dynamic evolution can be described by the Gross-Pitaevskii (GP) equation, and that the interaction parameters can be experimentally controlled through techniques such as evaporative cooling and Feshbach resonance \cite{evaporative cooling,BECbook, Feshbach,Feshbach1,Feshbach2,Feshbach3}. The initial states of BECs can be readily controlled by quantum state engineering technology \cite{engineering1, engineering2, engineering3, engineering4}. These above provides convenience for studying the nonlinear properties of BECs in experiments. Many nonlinear phenomena such as rogue waves (RWs) \cite{BECRW1, BECRW2, BECRW3, BECRW4, BECRW5}, solitons \cite{soliton1,soliton2,soliton3,soliton4} and breathers \cite{breather1,breather2} have been detected in BECs and their properties have been extensively studied.

     The term RW is originally coined in the field of oceanography to describe large, spontaneous, and unexpected water waves that can cause massive damage to navigation and offshore oil exploration \cite{Ocean1, Ocean2ML, Ocean3, Ocean4, Ocean5}. The Peregrine solution, which localizes both in space and time for the nonlinear Schrödinger equation (NLSE), represents a significant advancement in the study of RWs \cite{Peregrine}. The extensive variety of RW patterns, which are subjected to analytical investigation and classification within the context of the NLSE, offers the potential to provide insights that can be employed in the prediction of subsequent RW events based on the analysis of earlier wave forms \cite{patterns1, patterns2, patterns3, patterns4, patterns5}. Modulation instability (MI) is recognized as the primary mechanism for the generation of RWs \cite{MI1, MI2, MI3, MI4, MI5, MI6}. Moreover, stability analysis of RWs offers numerous possibilities for controlling the behavior of these nonlinear waves in a robust manner \cite{robustness1, robustness2}. RWs have also been subjected to experimental studies in a multitude of physical systems due to their profound physical implications, including water tanks \cite{watertank1, watertank2, watertank3}, non-linear optical systems \cite{opticalfiber1, opticalfiber2, opticalfiber3}, plasma \cite{plasma1}, and red blood cell suspensions \cite{RBC}. However, these studies necessitate the utilization of intricate initial perturbations and lack convenient control over RWs, particularly for the generation of high-order RWs. BEC is not subject to these limitations and is regarded as an appropriate physical platform for investigating RWs in a laboratory setting.
    
     There are two fundamental reasons for the great interest in generating RWs in numerical studies. Firstly, this provides the opportunity for in-depth studies of their properties and the testing of the applicability of mathematical models developed for their descriptions, which would be unfeasible in natural conditions. Secondly, it can provide guidance for the generation of RWs in experiments. 
     The results of numerical computation research indicate that first-order RWs can be excited and controlled to appear at fixed times and locations in BECs with attractive interactions under wide Gaussian initial conditions \cite{Gaussian, Gaussian2}. Subsequently, P. Engels and his collaborators use a two-component repulsive BEC with a highly asymmetric particle number, in which the effective interaction of the minority component is attractive. Using axial wide Gaussian initial conditions and the resulting modulation instability, they experimentally realize the Peregrine soliton, which is the prototype of the RW \cite{BECRW4}. This shows the guiding role of numerical simulations in experiments. Another recent numerical study demonstrated that when the two initial Gaussian wave packets are adjusted to appropriate offsets, the collision of two first-order RWs can generate a second-order RW \cite{collision}. The ``collision" mentioned here refers to the coincidence of the core structures generated when the two RWs evolve to their peak values, rather than the process of the nonlinear superposition evolution in the initial stage of the two wave packets. We will follow this concept of RWs collisions. However, when we reproduce the numerical results in the literature \cite{collision}, we are surprised to find that the collision of two first-order RWs is incapable of producing a second-order RW when the intensity of the interspecific attractive interaction is relatively weaker than that of the intraspecific interaction. We note a study of the collisions of two symmetric quantum droplets, which suggests that as the incident momentum increases, the colliding droplets may undergo a splitting process that results in the formation of two or even more droplets \cite{droplets}. What impact does the incident momentum have on the collision of RWs? In light of this, we introduce incident momentum and investigate its potential to facilitate the generation of high-order RWs.
     
     In this paper, we numerically solve the coupled two-component GP equations with two Gaussian wave packets as initial conditions, each having opposite incident momentum. It is demonstrated that, when appropriate incident momentum is employed, second-order RWs can be successfully generated in the case of weaker interspecies interactions compared to intraspecific interactions. Furthermore, we search the parameter plane in terms of incident momentum and interspecies interactions as coordinates and delineate the second-order RWs region. 

     This article is organized as follows. In sec. II, we provide a brief introduction to the formulation of two-component BECs in the mean field approximation, and present the experimental settings corresponding to the model parameters used in our subsequent numerical calculations. In sec. III, we present the promotion effect of appropriate incident momentum on the generation of second-order RWs in the case of weaker interspecies interactions compared to intraspecific interactions. Subsequently, we delineate the diverse ranges of incident momenta that can facilitate the generation of second-order RWs in varying interspecies interactions. We employ machine learning to identify the optimal initial offset that can generate second-order RWs for each parameter combination of interspecific interaction and incident momentum, and analyze the qualitative relationships among the above three parameters. In sec. IV, combined with our numerical findings, we explore the possible quantitative relationship between high-order RWs and the MI. We analyze the evolution of each component of the collision process under different incident momentum and point out the role of incident momentum in the generation of second-order RWs. Finally, the results are summarized in sec. V.
	
	\section{Formulation}\label{form}

    We begin with the two-component BECs that can be accurately described by the GP equation in the mean-field approximation \cite{BECbook}:
        \begin{equation}
        i \hbar \frac{\partial \Psi_{j}}{\partial t}=\left[\hat{H}_{j}+G_{j j}\left|\Psi_{j}\right|^{2}+G_{j, 3-j}\left|\Psi_{3-j}\right|^{2}\right] \Psi_{j},
        \end{equation}
    where $\Psi_{j} (j=1,2) $ is the macroscopic wave function of the condensates, and $\hat{H}_{j}=-\hbar^{2}\nabla^{2}/(2m)+m\omega_{\perp}^{2}(x^{2}+y^{2})/2$. $m$ is the mass of the atoms of each component, $\omega_{\perp}$ is the radial trap frequencies. $G_{j j}=4 \pi \hbar^{2} a_{j j} / m$ is the diagonal coupling coefficient, which represents the interaction between the same components, where $a_{jj}$ is the s-wave scattering length. $G_{12}=G_{21}=4\pi\hbar^{2}a_{12}\,/\,m$ is the non-diagonal coupling coefficient, which represents the interaction between different components.

    To simplify the calculations, we use the dimensionless form and focus on quasi-one-dimensional BECs. The BEC is trapped in the potential well and forms a cigar-shaped configuration. The radial motion of the condensate is then frozen. We can write the wave function as $\Psi_{j}(r,t)=\phi_{j}(x,y)\psi_{j}(z,t)$, where $\phi_{j}=\pi^{-1/2}e^{-(x^{2}+y^{2})}$ is the ground state wave function of the radial resonator potential, and $\psi_{j}(z,t)$ is the axial wave function. In the regime of weak interatomic coupling, selecting a Gaussian shape for the condensate in the radial direction is highly reasonable. This is because the precise ground state of the linear Schrödinger equation featuring a harmonic potential is indeed a Gaussian \cite{RaGs1}. Furthermore, when it comes to describing the collective dynamics of BECs, it has been previously demonstrated that the variational method relying on Gaussian trial functions yields reliable results even in the large condensate number limit \cite{RaGs2}, and the results are in good agreement with the experiment in BEC \cite{RaGs3}. After integration over the transverse coordinates, the coupled GP equations are in a dimensionless and efficient one-dimensional form:
        \begin{equation}
        \begin{split}
            i \frac{\partial \psi_{1}}{\partial t}=\left[-\frac{1}{2} \frac{\partial^2}{\partial{z^2}}+\mathsf{g}_{11}\left|\psi_{1}\right|^{2}+\mathsf{g}_{12}\left|\psi_{2}\right|^{2}\right] \psi_{1}, \\
            i \frac{\partial \psi_{2}}{\partial t}=\left[-\frac{1}{2} \frac{\partial^2}{\partial{z^2}}+\mathsf{g}_{22}\left|\psi_{2}\right|^{2}+\mathsf{g}_{21}\left|\psi_{1}\right|^{2}\right] \psi_{2},
        \label{eq2}%
        \end{split}
        \end{equation}    
    where $\mathsf{g}_{11}=2Na_{11}/a_{\perp}$ , $\mathsf{g}_{22}=2Na_{22}/a_{\perp}$ , $\mathsf{g}_{12}=\mathsf{g}_{21}=2Na_{12}/a_{\perp}$ are dimensionless interaction parameters. $N$ is the total number of particles. The number of particles in each of the two components is conserved and both are equal to $N/2$. $a_{\perp}=\sqrt{\hbar/(m\omega_{\perp})}$  is the width of the radial harmonic oscillator. We use the following dimensionless units: $\omega_{\perp}^{-1}$ for time, $a_{\perp}$ for length, and $\sqrt{N/{a_{\perp}}^3}$ for the wave function. In addition, the dimensionless complex-valued macroscopic wave function is defined as $\Psi:=\Psi(z,t)=(\psi_{1}(z,t),\psi_{2}(z,t))^{T}$. It is normalized as 
	\begin{equation}
		\label{eq3}
		\int \Vert \Psi \Vert_2^{2}dz=\int {n(z,t)}dz=1,
	\end{equation}
	where $n(z,t)$ is the evolution of the total particle density, which is described as
        \begin{equation}
            n(z,t)=\left|\psi_{1}(z, t)\right|^{2}+\left|\psi_{2}(z, t)\right|^{2}.
        \end{equation}	
	
    We use the Gaussian function with incident momentum as the initial condition for each component, which are denoted as
        \begin{equation}
            \psi_{1}(z,0)=C_{1}\exp\left(-{\frac{(z-\delta)^{2}}{2\sigma^{2}}}\right)\!e^{-ikz},
        \label{eq5}%
        \end{equation}
        \begin{equation}
            \psi_{2}(z,0)=C_{2}\exp\left(-{\frac{(z+\delta)^{2}}{2\sigma^{2}}}\right)\!e^{ikz},
        \label{eq6}%
        \end{equation}
    where $\sigma$ is the width of the Gaussian wave packet, $\delta$ is the offset of $\psi_{1}(z,0)$ and $\psi_{2}(z,0)$ from the center position, $k$ represents the incident momentum of the wave packet, $C_{1} = C_{2}$ represent the normalization coefficients. The initial condition satisfies Eq. (\ref{eq3}).

\begin{figure}
		\includegraphics[width=0.45\textwidth]{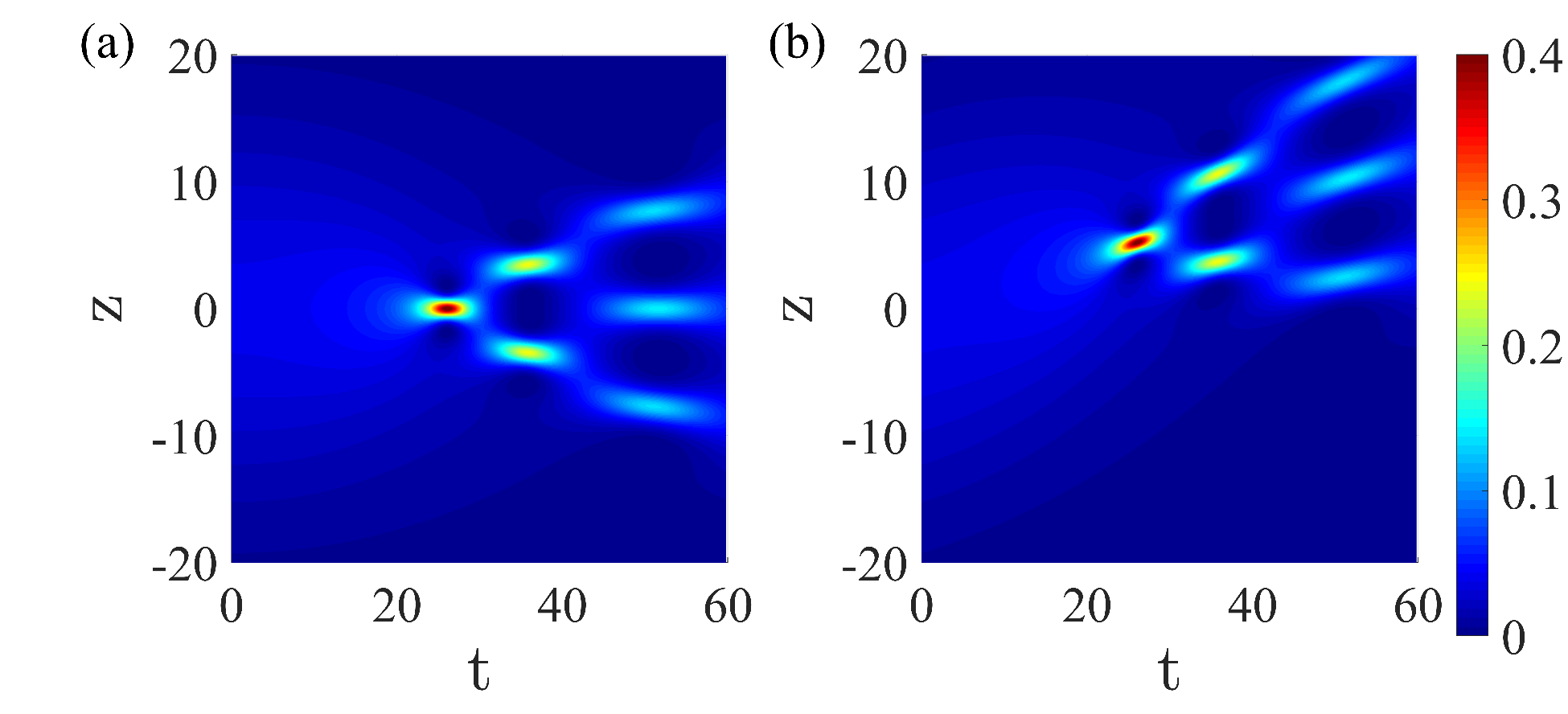}  \centering
		\caption{BEC evolution dynamics over time when $\sigma=10\sqrt2$, $\delta=0$, $\mathsf{g}_{11} = \mathsf{g}_{22}=-6$, $\mathsf{g}_{12} = \mathsf{g}_{21}=-3$, $C_{1}=C_{2}$. Colors represent total particle number density. The propagation mode of first-order RW shows a layered pattern. (a) shows the evolution when the initial momentum $k=0$, and (b) represents the situation when the initial momentum is 0.2 in the same direction.}
		\label{Fig1}
	\end{figure}

    In fact, first-order RWs can be observed under attractive interactions and larger wave packet widths \cite{Gaussian}. When $\delta$ is small, the evolution diagrams of the two components overlap each other, showing a structure similar to first-order RW. When $k=0$, as shown in Fig. \ref{Fig1}(a), the propagation mode of the first-order RW shows layered propagation as time evolves, the first-order RW becomes two first-order RWs as it propagates over time, then two first-order RWs turn into three first-order RWs and so on, this mode is also called the ``Christmas tree" mode. When the incident momentum in the same direction is set, as shown in Fig. \ref{Fig1}(b), it maintains the ``Christmas tree" mode, but propagates in one direction. 
    
    In this article we examine the collision between first-order RWs evolved from $\psi_{1}(z,0)$ and $\psi_{2}(z,0)$. When the two components are separated by appropriate offset $\delta$ at the initial time, each component generates a first-order RW and propagates in a ``Christmas tree" mode. They propagate diagonally and collide due to the presence of interspecific attractive interactions and initial momentum in opposite directions. It has been demonstrated that when the peaks of two first-order RWs coincide in both position and time, a second-order RW can be initiated \cite{collision}, and the peak of the second-order RW produced at this time is the largest. However, we find that the collision of two first-order RWs is incapable of producing a second-order RW in the case of weaker interspecies interactions compared to intraspecific interactions. To address this issue, we employ a numerical solution of the coupled GP Eq. (\ref{eq2}), and use Gaussian functions with incident momentum Eq. [(\ref{eq5})-(\ref{eq6})] as initial conditions. The boundary conditions are $\psi_{1}(z = \pm 50) = \psi_{2}(z = \pm 50) = 0$. Obviously, $\sigma$, $\mathsf{g}_{11}$, $\mathsf{g}_{22}$, $\mathsf{g}_{12}$, $\mathsf{g}_{21}$, $\delta$ and $k$ are parameters used to control our system. In all subsequent calculations in the main body of this paper, we set $\sigma$ = $10\sqrt2$ and $\mathsf{g}_{11} = \mathsf{g}_{22}=-6$ to easily generate RWs, assume $\mathsf{g}_{12} = \mathsf{g}_{21}$ for simplicity.
    
    Taking the two hyperfine states of \(^{87}\text{Rb}\) atoms as an example, the total number of atoms is adjusted to \(N \approx 10^{4}\). Each hyperfine state is occupied by half of the atoms, and the BEC is confined in a highly elongated harmonic trap with a radial frequency \(\omega_{\perp} = 2\pi \times 200 \, \text{Hz}\) and an axial frequency \(\omega_{z} = 2\pi \, \text{Hz}\). The 200:1 aspect ratio of the optical trap ensures effective 1D dynamics. The numerical results show that the effect of the trap frequency aspect ratio \(\lambda \rightarrow 0\) on the evolution of the RW is minimal and does not affect our conclusions. Therefore, we omit the potential term in the numerical simulations of the GP Eq. (\ref{eq2}). The radial length \(a_{\perp} = \sqrt{\hbar / m \omega_{\perp}} \approx 0.71 \, \mu\text{m}\). The hyperfine states can be controlled via Feshbach resonance to adjust the intraspecies scattering lengths \(a_{jj}\) and the interspecies scattering length \(a_{12}\). When adjusting \(a_{11} = a_{22} = -4a_{0}\) (where \(a_{0}\) is the Bohr radius), \(\mathsf{g}_{11} = \mathsf{g}_{22} = 2N a_{11} / a_{\perp} \approx -6\). Similarly, when adjusting \(a_{12} = -2.3a_{0}\), \(\mathsf{g}_{12} = 2N a_{12} / a_{\perp} \approx -3.5\), which corresponds to the situation where the interspecific interaction strength is weaker compared to the intraspecific interaction strength.

    Therefore, the only remaining undetermined parameters are $\mathsf{g}_{12}$, $k$ and $\delta$. For each set of $\mathsf{g}_{12}$ and $k$, we adjust $\delta$ to make sure that two first-order RWs meet and the peak amplitude of the collision is the largest.  

	\begin{figure*}
		\includegraphics[width=0.85\textwidth]{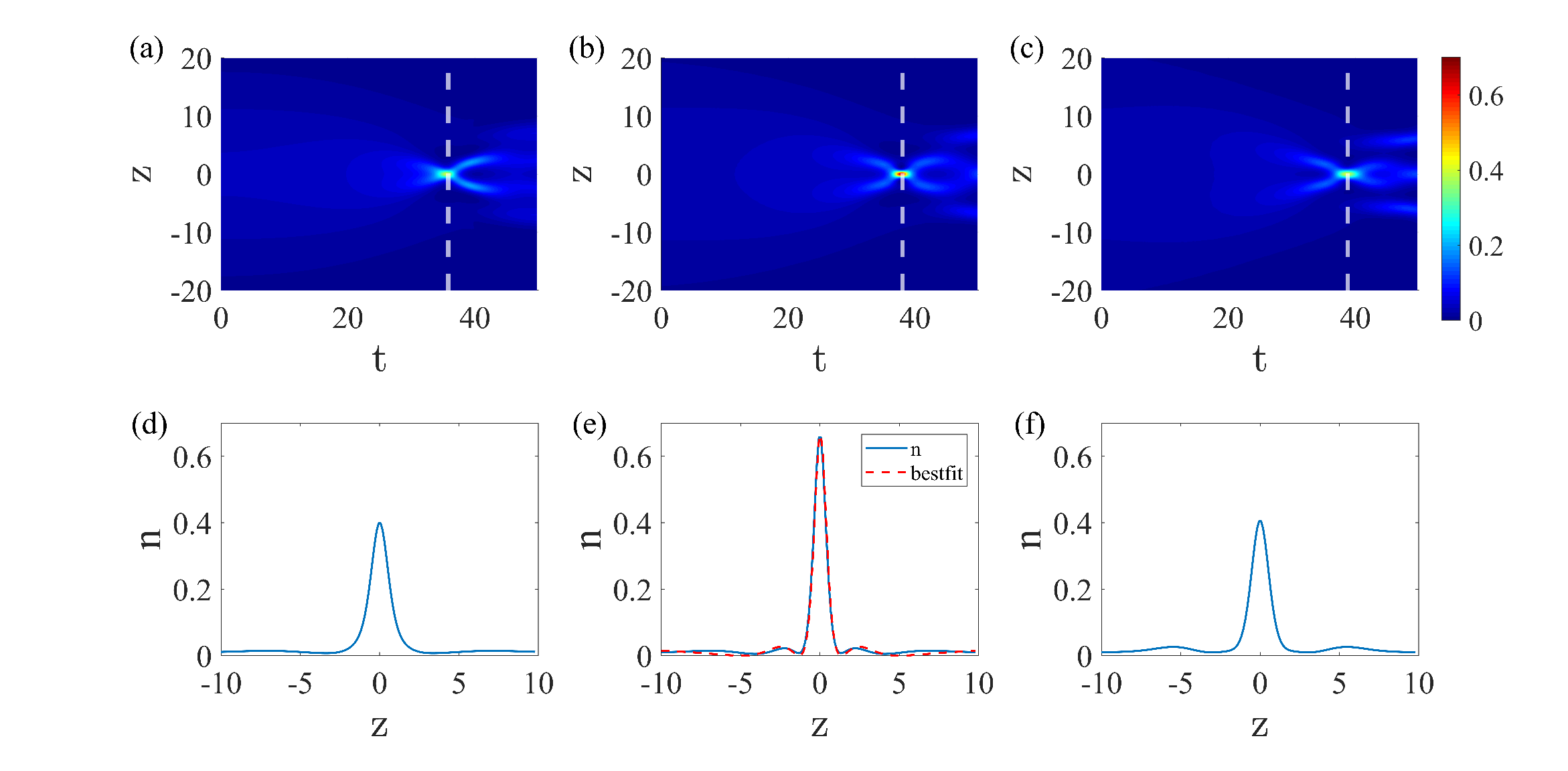}  \centering
		\caption{RWs collision with parameter $\mathsf{g}_{12} = -3.5$ and $\mathsf{g}_{11} = \mathsf{g}_{22} = -6$. The first row [(a)-(c)] shows the space-time evolution of the total particle number density $n(z,t)$ under different incident momentum $k$, each column $k$ is set to 0, 0.1 and 0.2 respectively, and $\delta$ is set to 5.4, 7.9 and 10.1. The second row [(d)-(f)] respectively shows the detailed structure of the maximum amplitude moment corresponding to the first row. (e) also shows the fitting plot of second-order RWs generation by collision (blue line) to second-order RW analytic solution (red dotted line).}
		\label{Fig2}
	\end{figure*}

	\begin{figure}
		\includegraphics[width=0.45\textwidth]{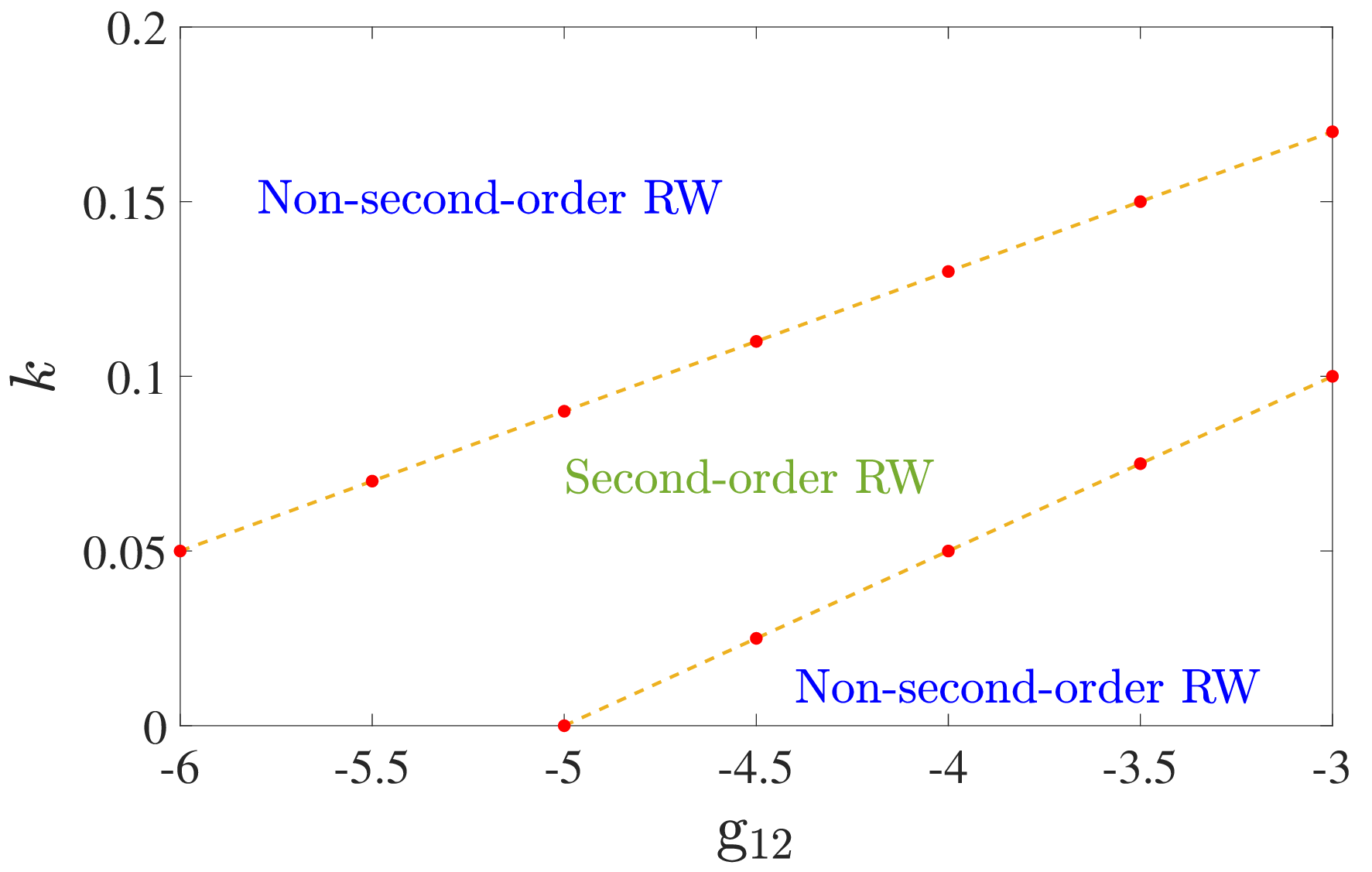}  \centering
		\caption{The parameter plane of the collision dynamics of two RWs for $\mathsf{g}_{11} = \mathsf{g}_{22} = -6$. The parameter regions in the upper-left and lower-right corners cannot produce second-order RW, and the region in the center can produce second-order RW. The boundary of the region where the second-order RWs are generated is determined by the number of local minimum points in the spatial distribution at the peak $n$ moment.}
		\label{Fig3}
	\end{figure}

	\begin{figure*}
		\includegraphics[width=0.85\textwidth]{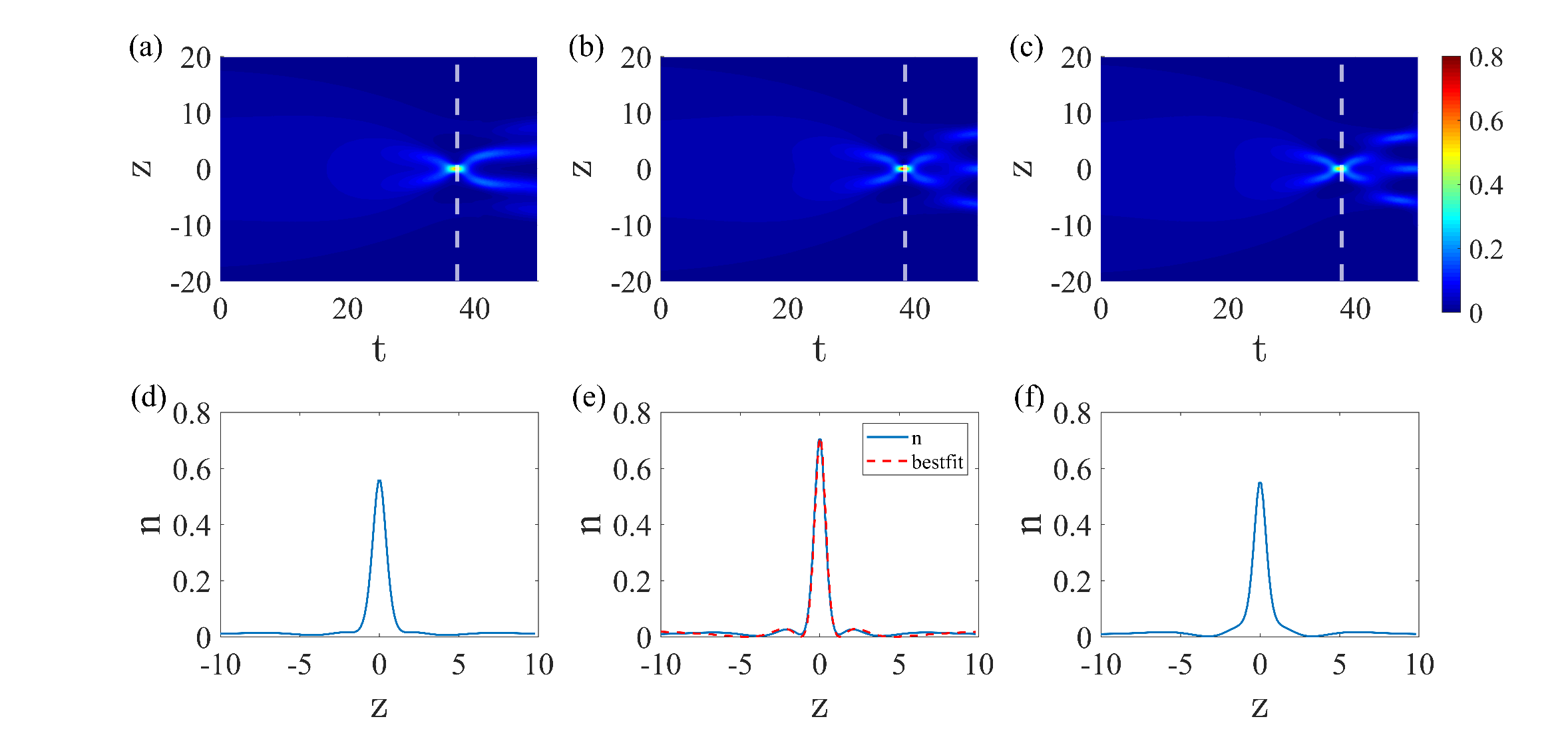} \centering
		\caption{RWs collisions mediated by two-component BEC interactions based on fixed momentum $k=0.075$ and $\mathsf{g}_{11} = \mathsf{g}_{22} = -6$. The first row [(a)-(c)] shows the spatiotemporal evolution of the total particle number density $n(z,t)$ under different interspecies interactions $\mathsf{g}_{12}$, each column sets $\mathsf{g}_{12}$ to -3, -4.5 and -6. The second row [(d)-(f)] respectively shows the detailed structure of the maximum amplitude moment corresponding to the first row. (e) also shows the fitting plot of second-order RWs generation by collision (blue line) to second-order RW analytic solution (red dotted line).}
		\label{Fig4}
	\end{figure*} 
 
	\begin{figure*}
		\includegraphics[width=0.95\textwidth]{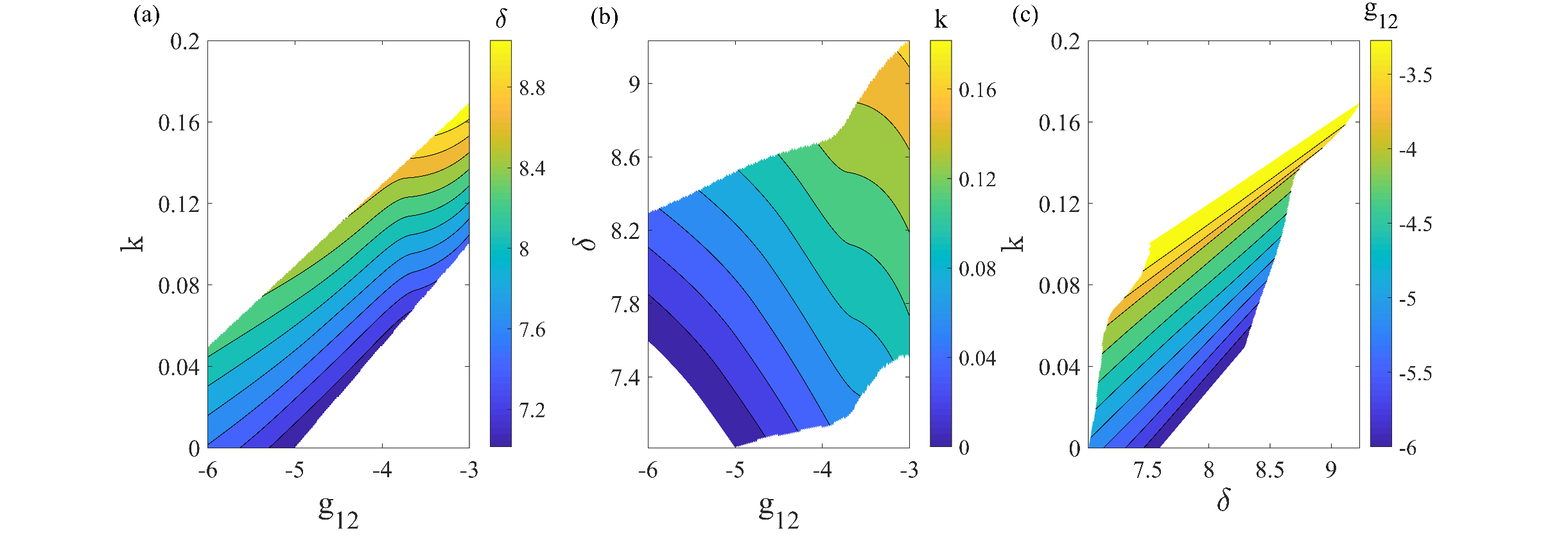}  \centering
		\caption{Three views of the parameters of the second-order RWs region.The color bar of [(a) - (c)] represents the parameters $\delta$, $k$, and $\mathsf{g}_{12}$ respectively, and the black lines are contours.}
		\label{Fig5}
	\end{figure*}
	
	\begin{figure*}
		\includegraphics[width=0.95\textwidth]{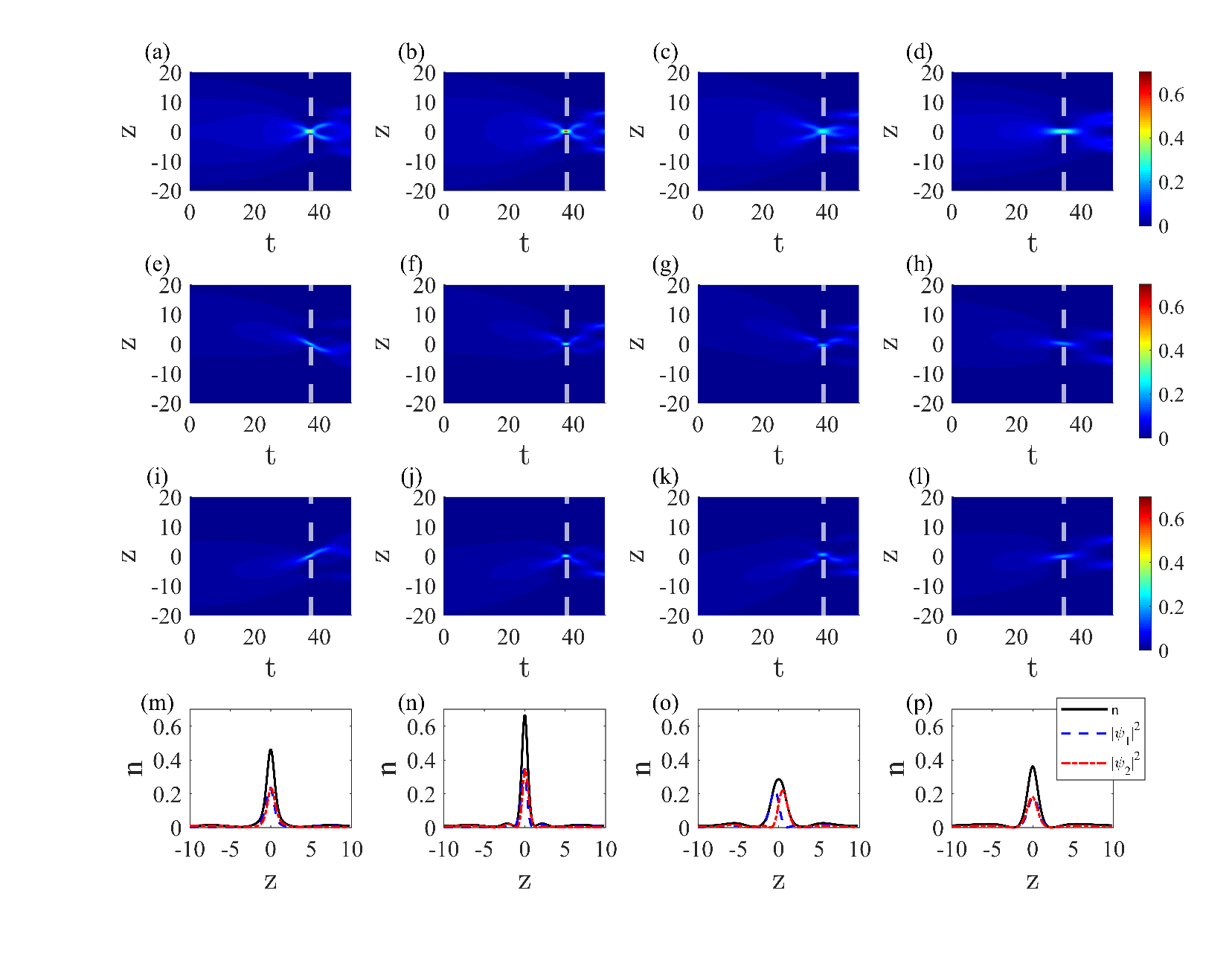}  \centering
            \caption{Details of the evolution of each component of the collision process for different parameters. The first row [(a)-(d)] shows the spatiotemporal evolution of the total particle number density $n(z,t)$, the second row [(e)-(h)] and the third row [(i)-(l)] respectively show the temporal and spatial evolution of each component $\left|\psi_{1}(z, t)\right|^{2}$ and $\left|\psi_{2}(z, t)\right|^{2}$ corresponding to the first row, the fourth row [(m) - (p)] shows the detailed structure of the maximum amplitude moment of the first three rows. The parameters in the first three columns are $\mathsf{g}_{11} = \mathsf{g}_{22} = -6$, $\mathsf{g}_{12} = -4$, $k =$ 0, 0.1 and 0.2 respectively, and the parameters in the fourth column are $\mathsf{g}_{11} = \mathsf{g}_{22} = -6$, $\mathsf{g}_{12} = 0$ and $k = 0.1$.  }
		\label{Fig8}
	\end{figure*}

    \section{Results}\label{resu}

    How should a second-order RW be defined? In the exact solution, when the amplitude of the RWs reaches its maximum value, the N-order RW exhibits 2N zero points distributed in space, with the maximum amplitude being $2N+1$ times the background amplitude \cite{exact solution}. The second-order RW has four troughs around the peak, with peaks five times as strong as the background, which was first observed in water wave tank experiments \cite{watertank2}. Since the second-order RW in BEC is expressed here in terms of the particle number density $n$, it can be deduced that the peak density of the second-order RW should be 25 times that of the background. In this paper, we consider a second-order RW to occur if the peak of RW is sufficiently high and surrounded by four troughs.

	\subsection{Generate second-order RWs under modulation of $k$}\label{subsectionA}

    In this section, we employ a numerical simulation to investigate the collision process under different incident momentum conditions when the interspecies interaction between the two bodies is relatively weak ($\mathsf{g}_{12}=-3.5$) compared to the fixed value of $\mathsf{g}_{11}=\mathsf{g}_{22}=-6$. In the absence of incident momentum ($k$=0), no matter how the offset is adjusted, no second-order RW is produced. As illustrated in Fig. \ref{Fig2}(a), the offset is adjusted to 5.4 for the most appropriate, the peak density of the wave packet is not high enough, and there are only two troughs around the peak. When an attempt is made to introduce a small incident momentum ($k=0.1$) and the appropriate offset is adjusted to $7.9$, as shown in Fig.~\ref{Fig2}(b), the peak density of the wave packet achieves $22.5$ times the initial wave height, and there are four troughs around the peak. In addition, we compare the particle density $n$ of numerical results with the analytical solution of the second-order RWs, and clearly the structure of the generated second-order RW is consistent with the structure of the exact solution \cite{exact solution}, as shown in Fig.~\ref{Fig2}(e). Consequently, the introduction of incident momentum successfully leads to the generation of second-order RW. 

    For the convenience of subsequent discussions, the value of \(\mathsf{g}_{11} = \mathsf{g}_{22}\) is fixed at -6 in the main body of this paper. When \(\mathsf{g}_{11} = \mathsf{g}_{22}\) takes other values, the conclusion remains the same. That is, when \(|\mathsf{g}_{12}|\) is relatively smaller than \(|\mathsf{g}_{11}|\), without introducing the incident momentum \(k\), it is impossible to generate the second-order RW no matter how the offset is adjusted. However, when an appropriate incident momentum is introduced, the second-order RW can be generated by adjusting the offset. An example with \(\mathsf{g}_{11}=\mathsf{g}_{22}=-10\) is presented in the Appendix. 
    
    Yet not any arbitrary value of $k$ can promote the generation of second-order RWs. For example, if we increase $k$ to 0.2, though the offset is adjusted to the most appropriate value of $\delta=10.1$, as shown in Fig.~\ref{Fig2}(c), the collision also fails to produce a second-order RW. We explore alternative values of $k$ and discovered that the range of $k$ that allows the generation of second-order RWs is from $0.075$ to $0.15$ when $\mathsf{g}_{12}=-3.5$. It is evident that $k$ exerts a significant influence on the collision of RWs, appropriate $k$ can facilitate the generation of second-order RWs. This phenomenon is elucidated in the section of discussion.
	
	\subsection{Second-order RWs region parameterized by $\mathsf{g}_{12}$ and $k$}
	
    One of the advantages of BEC is the adjustability of interaction, the strength of $\mathsf{g}_{12}$ can also influence the dynamic evolution of RWs. In section A, it is demonstrated that for $\mathsf{g}_{12}=-3.5$, the $k$ that promotes the generation of second-order RWs has a range. So we scan the parameter plane to ascertain the range of $k$ under different $\mathsf{g}_{12}$. As plotted in Fig. \ref{Fig3}, the consequences of the collision are classified into two categories: second-order RWs region and non-second-order RWs region. It can be observed that second-order RWs cannot be generated in the absence of $k$ when $\mid\mathsf{g}_{12}\mid < 5$. We examined the cases $\mathsf{g}_{12}=-3, -4.5$ and $-6$ at fixed $k=0.075$. When $\mathsf{g}_{12}=-4.5$, the parameter falls within the second-order RW region which can produce second-order RWs. As illustrated in Fig. \ref{Fig4}(b), the peak density of the wave packet reaches 24.2 times the initial wave height, and there are four troughs around the peak. When $\mathsf{g}_{12}=-3$ or $-6$, the parameter falls outside the second-order RW region. As illustrated in Fig. \ref{Fig4}(a) and 4(c), second-order RWs cannot be generated, regardless of the offset adjustment. Since these two parameter points are close to the boundary of the second-order RW region, they also produce high peaks, but there are only two troughs around the peak. Furthermore, in the case where $\mid\mathsf{g}_{12}\mid$ is too small, the nonlinear attractive interactions are too weak to produce second-order RWs no matter how $k$ is adjusted. The trend of the second-order RWs region in Fig. \ref{Fig3} reveals that when $\mid\mathsf{g}_{12}\mid$ is excessive, the collision of RWs is also unable to generate second-order RWs. This is also corroborated by the numerical results.
    
        \subsection{Searching for control parameters using machine learning}  

    As previously stated, for each set of parameters comprising $\mathsf{g}_{12}$ and $k$, we adjust the offset $\delta$ in order to ensure that two first-order RWs meet and that the peak amplitude of the collision is the largest. It is obvious that identifying $\delta$ within the entirety of the second-order RW region is a time-consuming endeavor, so $\delta$ is not provided in Fig. \ref{Fig3}. The application of machine learning to research on RW has yielded promising results \cite{ML1, ML2, Ocean2ML, ML3, ML4, ML5}, and it is well suited to solve our problem above. In order to identify the optimal $\delta$ for the entire second-order RWs region, we employ a deep learning neural network (DNN). The inputs to the network are the parameters $\mathsf{g}_{12}$ and $k$, and the output is $\delta$. A total of 70 parameter combinations of  $\mathsf{g}_{12}$, $k$ and $\delta$ are collected within the second-order RWs region, and half of the parameter combinations are used for training purposes, while the remaining half are employed for testing. We adopt the following mean squared error (MSE) loss function :
        \begin{equation}
        L_{NN}=\frac{1}{N_{\mathrm{data}}}\sum_{n = 1}^{N_{\mathrm{data}}}\left(\delta_{n}^{NN}(\mathbf{g_{12}},\mathbf{k})-\delta_{n}\right)^{2},
        \end{equation}
    where $N_{data}$ denotes the total number of training data points; and $\delta_{n}^{NN}$ and $\delta_{n}$ are the prediction of DNN and the true data, respectively. After the training is completed, the mean relative error of the test set is $0.0007$, indicating that the training results are effective. Three views of the second-order RWs region are created, as illustrated in Fig. \ref{Fig5}. The contour lines reveal that when the value of $\delta$ remains constant, the value of $k$ decreases with an increase in $\mid\mathsf{g}_{12}\mid$, as shown in Fig.~\ref{Fig5}(a). Conversely, when $k$ remains constant, $\delta$ increases with an increase in $\mid\mathsf{g}_{12}\mid$, as shown in Fig.~\ref{Fig5}(b). Furthermore, when $\mathsf{g}_{12}$ remains constant, $\delta$ increases with an increase in $k$, as shown in Fig.~\ref{Fig5}(c). It is worth noting that the qualitative relationships of the above three parameters are invariant even outside the second-order RWs region for two RWs to collide and produce the largest peak.
	
	\section{Discussion } 
    Why do second-order RWs appear only in combinations of the parameters $\mathsf{g}_{12}$, $k$ and $\delta$ within the boundary of the second-order RW region? Perhaps this can be understood from the quantitative correlation between high-order RWs and MI. The exponential growth of perturbations on the plane wave background driven by the MI creates conditions for the formation of nonlinear structures such as RWs and breathers \cite{MI4}. When the perturbation frequency approaches zero within the fundamental MI band, the RWs will be excited \cite{MI1}. Apart from the eye-shaped Peregrine RW, fundamental RWs have anti-eye-shaped and four-petal structure. All are excited when the perturbation frequency approaches zero in the fundamental band. What parameters determine the difference of RW structure?  L.-C. Zhao, et al. show that the mode type of the fundamental RW can be directly and quantitatively given by the range of a discriminant, which is related to the real and imaginary parts of the perturbation wavenumber and the background frequency \cite{MI6}. Then, according to the excitation conditions of fundamental RWs with different structures, they numerically construct a very simple form initial state (which can deviate from the initial state of the exact solution) that meets the excitation condition, and the corresponding  structure fundamental RW is excited. 

    High-order RWs are essentially the nonlinear superposition of multiple fundamental RWs of the same type and should have a more precise correspondence with MI. The parameter boundaries in Fig.~\ref{Fig3} reflect the key thresholds of the instability driven by the MI to support the formation of second-order RWs, especially the introduction of an additional perturbed incident momentum when the interspecific interaction is relatively weak compared to the intraspecific interaction. However, the quantitative relationship between high-order RWs and MI remains an unsolved frontier problem and is not the subject of this paper. Our work numerically verifies the hypothesis that high-order RWs are related to the characteristics of the MI, laying the foundation for future research to derive the analytical relationship between the MI frequency spectrum and the hierarchy of RWs.  
    
    In addition, to better observe how different incident momenta affect the generation of second-order RWs, we show the details of the evolution of each component for different parameters in Fig. \ref{Fig8}. The first row [(a)-(d)] of Fig. \ref{Fig8} shows the spatiotemporal evolution of the total particle number density $n(z,t)$, the second row [(e)-(h)] and the third row [(i)-(l)] respectively show the temporal and spatial evolution of each component $\left|\psi_{1}(z, t)\right|^{2}$ and $\left|\psi_{2}(z, t)\right|^{2}$, the fourth row [(m) - (p)] shows the detailed structure of the maximum amplitude moment of the first three rows. When $k=0$ and $\mathsf{g}_{12}=-4$, each component shows a structure similar to first-order RW, as shown in Fig. \ref{Fig8}(e) and 6(i), and the collision cannot produce a second-order RW, as shown in Fig. \ref{Fig8}(m). Second-order RW cannot be formed for the following reasons, when $\mathsf{g}_{12}=-4$ and $k=0$ are both relatively small, to make the two RWs collide, the offset should be small. The distance between the two wave packets is so close that there is not enough space for each wave packet to evolve. Moreover, our recent research on three-component BEC systems has also demonstrated that the offset plays a significant role in RWs collisions \cite{TTL2}. When we set $k=0.1$ and keep $\mathsf{g}_{12}=-4$ constant, the structure of each component is similar to the second-order RW, as shown in Fig. \ref {Fig8}(f) and 6(j). Each structure is asymmetric, but the sum of the two structures is symmetric, ultimately resulting in the formation of a second-order RW, as shown in Fig. \ref{Fig8}(n). This phenomenon can be attributed to the linear modulation of the incident momentum, as illustrated in Fig. \ref{Fig1}. This modulation maintains the requisite distance between the two wave packets throughout the evolution, ultimately leading to the emergence of a second-order RW.

    In Gaussian initial conditions, intraspecies interactions result in the generation of first-order RW, while intraspecies and interspecies interactions together result in second-order RW. This is analogous to the collision of solitons in a two-component BEC system. The attractive intraspecies interactions are crucial for the formation of soliton-like localized matter waves, while the attractive interspecies interactions have been observed to broaden the wave packets. Together, these two complementary interactions permit the wave packet to move as a breather \cite{interaction}. RW is a nonlinear phenomenon, and second-order RW cannot be generated under the Gaussian initial setup if there is only linear modulation of the incident momentum and no interspecies interactions. For instance, we set $\mathsf{g}_{12}=0$ and $k=0.1$, each component produces a first-order RW individually, as shown in Fig. \ref{Fig8}(h) and 6(l). However, the total evolution is only a simple superposition of each component, and there is no second-order RW produced, as shown in Fig. \ref{Fig8}(p). Furthermore, excessive linear modulation is deleterious to the generation of second-order RW. For example, we set an inappropriate $k=0.2$, and keep $\mathsf{g}_{12}=-4$ constant, although each component also shows a structure similar to second-order RW, these two structures collide at a big angle, as shown in Fig. \ref{Fig8}(g) and 6(k). When the two RWs collide and produce the largest peaks, the maxima of each component do not coincide, and no second-order RW is produced, as shown in Fig. \ref{Fig8}(o). This is due to the fact that the nonlinear attractive interaction $\mathsf{g}_{12}$ is unable to play a dominant role in the evolution of RWs when subjected to the influence of a big linear modulation of $k$. This ultimately results in the disruption of the generation of second-order RWs.
    
    Moreover, in two-component BEC systems, the introduction of appropriate phase factor $e^{ikz}$ can promote the generation of second-order RWs without a complicated interference effect. This is different from a single component system, the introduction of phase factors lead to complex interference \cite{phases,phases2}, which is not conducive to the generation of second-order RWs.
    
	\section{Conclusions}\label{concl}

    This paper examines the impact of incident momentum on the generation of second-order RWs in two-component BEC systems. The results demonstrate that appropriate incident momentum can facilitate the generation of second-order RWs in the case of weaker interspecies interactions compared to intraspecific interactions. This provides guidance for generating and controlling high-order RWs for related experimental studies. We determine the second-order RWs region parameterized by $\mathsf{g}_{12}$, $k$ and $\delta$, which reflect the key thresholds of the instability driven by the MI to support the formation of second-order RWs. Our work numerically verifies the hypothesis that high-order RWs are related to the characteristics of the MI, laying the foundation for future research to derive the analytical relationship between the MI frequency spectrum and the hierarchy of RWs.

	\begin{acknowledgments}
		
		This work was supported by the NSFC (Grant Nos. 11904309 and 11847096) and by the Natural Science Foundation of Hunan Province (Grant No.2020JJ5528).
		
	\end{acknowledgments}
    
        \begin{appendices}
        
        \renewcommand{\appendixname}{APPENDIX}
  
        \setcounter{secnumdepth}{0}

            \section*{APPENDIX: RW COLLISION WITH \(\mathsf{g}_{11}=\mathsf{g}_{22}=-10\)}
\begin{figure*}
    		\includegraphics[width=0.95\textwidth]{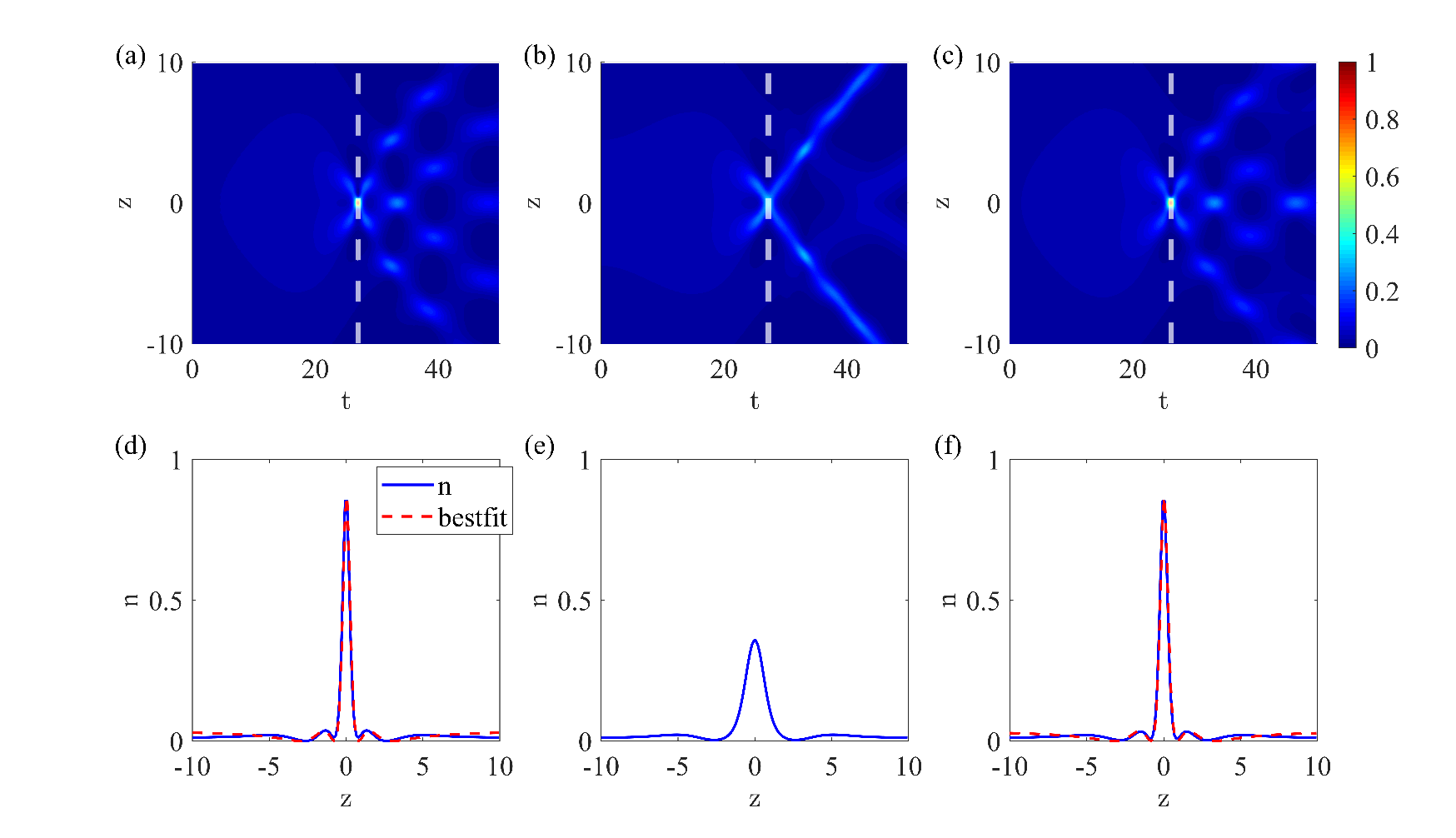}  \centering
                \caption{RWs collision for $\mathsf{g}_{11} = \mathsf{g}_{22} = -10$ and $\sigma=10\sqrt2$. The first row [(a)-(c)] shows the spatiotemporal evolution of the total particle number density $n(z,t)$, the second row [(d)-(f)] shows the detailed structure of the maximum amplitude moment of the first row. The parameters in the first column are $\mathsf{g}_{12} = -10$, $k =$ 0 and $\delta=7.2$. The parameters in the second column are $\mathsf{g}_{12} = -7$, $k =$ 0 and $\delta=6$. The parameters in the third column are $\mathsf{g}_{12} = -7$, $k = 0.1$ and $\delta=7.2$. (d) and (f) also show the fitting plot of second-order RWs generation by collision (blue line) to second-order RW analytic solution (red dotted line).}
    		\label{Fig7}
    	\end{figure*}

        Here we show the case of \(\mathsf{g}_{11}=-10\) mentioned in Sec. \ref{subsectionA}. When the interspecies interaction strength is equal to the intraspecies interaction strength, a second-order RW can be generated without introducing incident momentum, as shown in Fig. \ref{Fig7}(a). When the interspecies interaction strength is relatively weaker than the intraspecies interaction strength, a second-order RW can only be generated by introducing an appropriate incident momentum, as shown in Fig. \ref{Fig7}(b) and (c). 
            
        \end{appendices}

\end{document}